\documentclass[aps,prl,a4paper,reprint,nofootinbib,superscriptaddress]{revtex4-1}

\usepackage{hyperref}
\usepackage{amsmath}
\usepackage{amsfonts}
\usepackage{amssymb}
\usepackage{color}
\usepackage[utf8]{inputenc}
\usepackage{graphicx}
\usepackage{microtype}
\usepackage{siunitx}
\usepackage{soul}

\newcommand{\header}[1]{\textbf{#1:}}

\begin{document}

\begin{flushleft}
LAPTH-070/16
\end{flushleft}

\title{Searching for Primordial Black Holes in the radio and X-ray sky}
\date{\today}

\author{Daniele Gaggero}
\email{d.gaggero@uva.nl}
\affiliation{GRAPPA, University of Amsterdam, Science Park 904, 1098 XH Amsterdam, Netherlands}
\author{Gianfranco Bertone}
\affiliation{GRAPPA, University of Amsterdam, Science Park 904, 1098 XH Amsterdam, Netherlands}
\author{Francesca Calore}
\affiliation{GRAPPA, University of Amsterdam, Science Park 904, 1098 XH Amsterdam, Netherlands}
\affiliation{LAPTh, CNRS, 9 Chemin de Bellevue, 74941 Annecy-le-Vieux, France}
\author{Riley M. T. Connors}
\affiliation{GRAPPA, University of Amsterdam, Science Park 904, 1098 XH Amsterdam, Netherlands}
\affiliation{API, University of Amsterdam, Science Park 904, 1098 XH Amsterdam, Netherlands}
\author{Mark Lovell}
\affiliation{GRAPPA, University of Amsterdam, Science Park 904, 1098 XH Amsterdam, Netherlands}
\affiliation{MPIA, K\"onigstuhl 17, D-69117 Heidelberg, Germany}
\author{Sera Markoff}
\affiliation{GRAPPA, University of Amsterdam, Science Park 904, 1098 XH Amsterdam, Netherlands}
\affiliation{API, University of Amsterdam, Science Park 904, 1098 XH Amsterdam, Netherlands}
\author{Emma Storm}
\affiliation{GRAPPA, University of Amsterdam, Science Park 904, 1098 XH Amsterdam, Netherlands}

\begin{abstract}
We model the accretion of gas onto a population of massive {primordial} black holes in the Milky Way, 
and compare the predicted radio and X-ray emission with observational data. 
We show that under conservative assumptions on the accretion process, the possibility that 
${\cal O}(10) \, M_\odot$ primordial black holes can account for all of the dark matter in the Milky 
Way is excluded at $5\sigma$ by a comparison with a VLA radio catalog at $1.4$ GHz, and at 
{ $\simeq 40\sigma$  by a comparison with a Chandra X-ray catalog ($0.5 - 8$ keV)}. 
We argue that this method can be used to identify {such a} population of {primordial} black holes 
with more sensitive future radio and X-ray surveys.
\end{abstract}

\maketitle
\header{Introduction}
The first direct detection of a gravitational wave signal, announced by the LIGO collaboration 
earlier this year~\cite{Abbott:2016blz} demonstrated the existence of $\sim 30 M_\odot$ black 
holes {(BHs)}, {prompting the suggestion~\cite{Bird:2016dcv,Clesse:2016vqa} that these objects are {\it primordial} black 
holes (PBHs) that may account for all of the dark matter (DM)~\cite{Jungman:1995df,Bertone:2004pz,Bertone:2010at} in the Universe.}
The connection between PBHs and DM has been extensively studied in the past (see 
e.g.~\cite{Ivanov:1994pa,Khlopov:2008qy,Carr:2009jm,Blais:2002nd,Afshordi:2003zb,Frampton:2010sw}), 
and a number of constraints exist on the cosmic abundance of PBHs {over a very wide mass range} 
(see the discussion below, and e.g. Ref. \cite{Carr:2016drx} for a recent review). 

In this Letter, we consider for the first time, in the context of PBH searches, the X-ray and radio emission from the Galactic Ridge region produced by the accretion of interstellar gas onto a population  of ${\cal O}(10) \, M_\odot$ PBHs in the Milky Way. 
Given current estimates of the bulge mass \cite{Portail2015}, if PBHs constitute all of the DM, there should 
be ${\cal O}(10^9)$ such objects within $2$ kpc from the Galactic center (GC). 
Since the inner part of the bulge contains high gas densities~\cite{Ferriere2007}, a significant fraction would 
inevitably form an accretion disk and emit a broad-band spectrum of radiation.
We show (fig. \ref{fig:radioBound})     that radio and X-ray data in the Galactic Ridge region rule out, {at 5 and 40$\sigma$ respectively,} 
the possibility that PBHs constitute all of the DM in the Galaxy, even under conservative assumptions on the physics of accretion. 

{Our limits arise from a realistic modeling of the accretion process, based on the observational evidence for inefficient accretion in the Milky Way today~\cite{Perna:2003,Pellegrini:2005}, 
and corroborate, with a completely independent approach, the exclusion of {  massive PBHs as DM candidates.}

\begin{figure}[ht!]
	\includegraphics[width=\columnwidth]{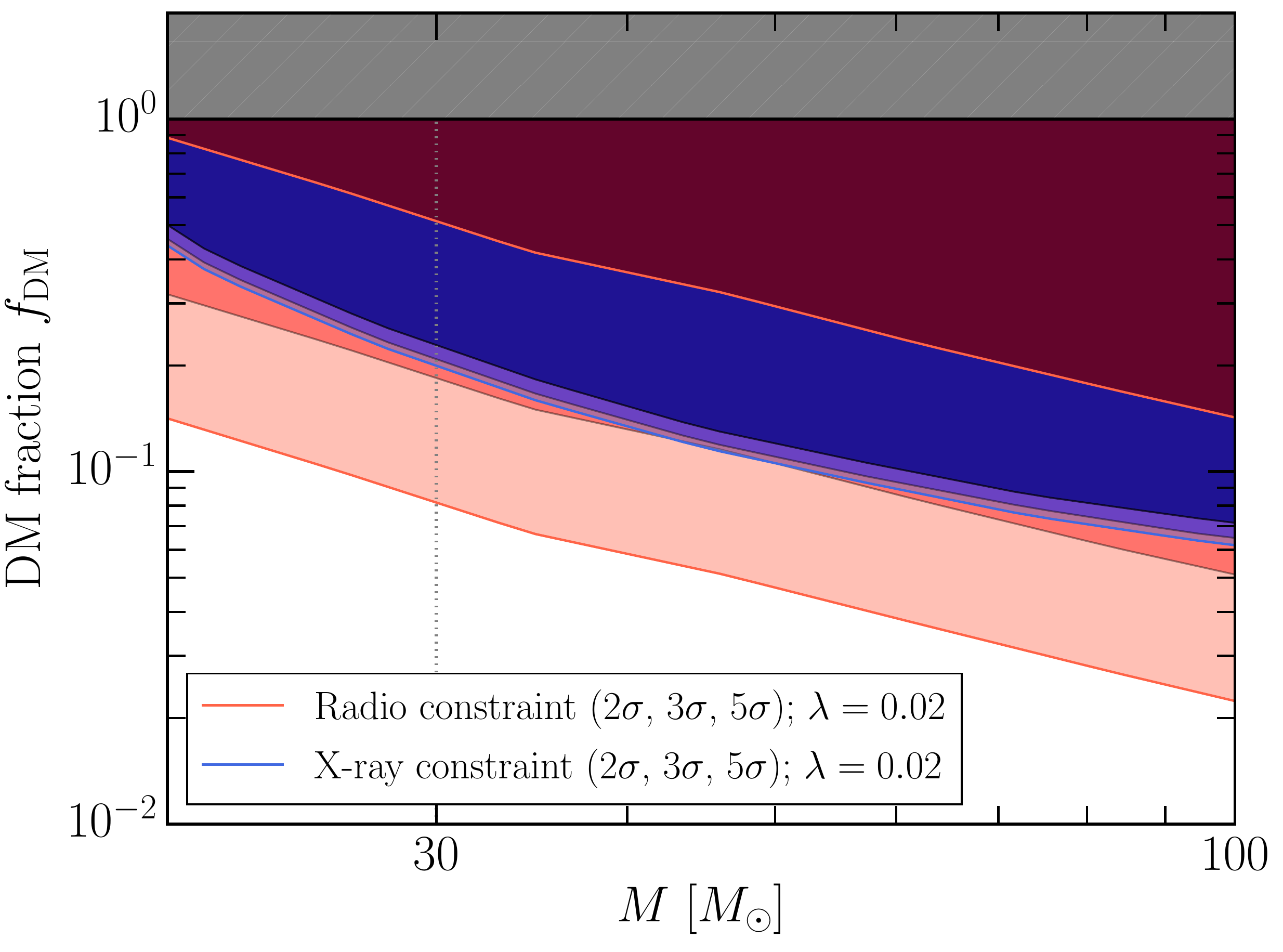}
    \caption{Upper limits on the fraction of DM in PBHs of a given mass $M$, arising from the non observation 
    	of bright X-ray (blue shaded regions) and radio (red) {BHs candidates} at the GC.    
	We assume a conservative value of $\lambda$, regulating the departure from Bondi accretion rate: $\lambda = 0.02$.
	The dotted grey line corresponds to 30$M_\odot$ PBH, the hatched grey region is unphysical ($f_{\rm DM} > 1$).
	}
    \label{fig:radioBound}
\end{figure}

\header{Accretion on black holes}
{ We should expect the accretion rates, $\dot{M}$, of a Galactic population of PBHs accreting from interstellar gas to be well below the Eddington 
limit $\dot{M}_{\rm Edd}$.} Even under the unrealistic assumption of Bondi-Hoyle-Lyttleton 
accretion~ \cite{Hoyle1939,Bondi1944}, and typical velocities as low as $\sim 10$ km/s, 
the accretion rate would definitely be  sub-Eddington: $\dot{M} \sim 10^{-5} \left( n_{\rm gas} / \rm cm^{-3} \right) \, \dot{M}_{\rm Edd}$. 

BHs accreting at { $\dot{M}<0.01~\dot{M}\rm_{Edd}\equiv\dot{M}_{crit}$} are radiatively inefficient, such that the luminosity scales non-linearly with $\dot{M}$ \cite{heinz2003}. The prevailing 
physical pictures adopted to explain the weak emission properties are advection-dominated accretion 
in which the gas cooling timescales greatly exceed the dynamical timescales \cite{Narayan:1994}, 
and mass loss from the disk or internal convective flows, such that the accretion rate itself has decreased once gas reaches
 the inner edge of the disk \cite{blandford:1999,quataert2000}. It is likely that both mechanisms are at play, 
 a view supported by both radio and X-ray constraints on the gas density around Sgr A*, the 
 supermassive BH at the center of the Galaxy, the least luminous accreting BH observed to date (in Eddingtion units), and thus a well-studied source from the point of view of weak accretion physics \cite{Bower2003,Marrone2007, wang2013}.
{  We compute the accretion rates and the radiative efficiencies of a Galactic population of PBHs in the low-efficiency limit, following the formalism presented in \cite{Maccarone2005MNRAS,Fender:2013}. We take into account the findings of previous studies regarding accretion of interstellar gas onto isolated black holes \cite{Fujita1998,Armitage1999,Agol2002}. }

We model the radiative efficiency $\eta$, defined by the relation for the bolometric luminosity $L_B = \eta \dot{M} c^2$, as $\eta = 0.1 \dot{M}/\dot{M}{\rm_{crit}}$ for $\dot{M}<\dot{M}\rm_{crit}$ (if we were to assume instead efficient accretion above the critical rate, $\dot{M}>\dot{M}\rm_{crit}$, then we would have a constant $\eta=0.1$). As already discussed, all our sources fall below this critical accretion rate, such that they are all inefficient accretors: this means the luminosity scales non-linearly with accretion rate, $L\propto\dot{M}^2$. 

We parameterize the accretion rate as 
 $\dot{M} = \lambda \dot{M}_{\rm Bondi}$, such that

\begin{equation}
\dot{M}=4\pi\lambda (G M_{BH})^2 \rho \left(v_{BH}^2+c_s^2\right)^{-3/2}
\label{eq:accretion}
\end{equation}

where $G$ is the gravitational constant, $v_{BH}$ is the velocity of the BH, and $c_s$ is the 
sound speed of the accreted gas, which is below 1 km/s in cold, dense environments. 

{  An important element that needs consideration is the temperature of the accreted gas due to radiative pre-heating \cite{Maccarone2005MNRAS}. Photoionising radiation will lead to an ionisation bubble surrounding the source, known as the Str\"{o}mgren sphere \cite{Stromgren1939}, with a characteristic radius, $R_{Str}$. In the following, we assume that the gas around the BH is fully ionized -- and therefore, we set $c_s = 10$ km/s -- if the timescale for the ionization of the Str\"{o}mgren sphere is shorter than the timescale associated with the incoming flux of fresh, unprocessed material.}

{ Regarding $\lambda$, we choose a reference value of $0.02$. Given the degeneracy between $\lambda$ and the angular momentum and temperature of the accreted gas, this value is consistent with isolated neutron  star population estimates and studies of active Galactic nuclei accretion~\cite{Perna:2003,Pellegrini:2005,wang2013}} 

{  This prescription is the same as that adopted by \cite{Fender:2013}; however, we consider $M_{BH}=30~M_{\odot}$, 
and rescale the value of $\dot{M}{\rm_{crit}}=0.01~\dot{M}_{Edd}$ used in that work across the full 10--100$~M_{\odot}$ mass range. }

We convert bolometric luminosity to X-ray luminosity via the approximate factor 
$L_{X} \simeq 0.3 \, L_B$ following \cite{Fender:2013}. 

Motivated by the results presented in \cite{Fender:2001} and by the discussion in \cite{Maccarone2005MNRAS, Fender:2013}, 
we assume the presence of a jet -- thus requiring a system with a surplus of angular momentum, {  or a dynamically important magnetic field combined with a spinning black hole} -- 
emitting radio waves in the GHz domain with an optically thick, almost flat spectrum, whilst the 
X-ray emission is non-thermally dominated, originating from optically thin regions closer to the 
BH. In order to convert the X-ray luminosity into a GHz radio flux, we adopt the universal empirical 
relation discussed e.g. in \cite{Plotkin:2012}, also known as the {\it fundamental plane (FP)}, which 
applies for a remarkably large class of compact objects of different masses, from X-ray binary systems to active Galactic nuclei. We calculate 
the X-ray luminosity in the 2--10 keV band in accordance with the FP, assuming a hard power-law X-ray spectrum with photon index $\alpha$, 
and a typical range for hard state X-ray binaries of 1.6--2.0 (see \cite{Hong:2016}). 
{ We extrapolate this power-law spectrum into the 0.5--8 keV and 10--40 keV bands in order to also make comparisons with Chandra and NuSTAR catalogs.} 
We then use the FP relation to calculate the 5 GHz 
radio flux from the 2--10 keV X-ray flux and assume a flat radio spectrum, such that $F_{\rm 5 \,GHz} = F_{\rm 1.4 \, GHz}$, allowing direct comparison with the 1.4 GHz source catalog from a VLA survey of the GC region.\\

\header{Primordial black hole population}
In order to derive a bound from X-ray and radio data, we set up a Monte Carlo simulation for 
each PBH mass, assuming a delta mass function.

We populate the Galaxy with PBHs following the Navarro-Frenk-White (NFW) distribution~\cite{Navarro:1995iw} 
 (other more conservative choices are discussed below). 
We implement the accurate 3D distribution of molecular, atomic, ionized gas in the inner bulge presented 
in~\citep{Ferriere2007}; that distribution includes a detailed model of the 3D structure of the Central Molecular 
Zone (CMZ), a 300 pc wide region characterized by large molecular gas density and centered on the GC, 
i.e. in the region where the largest density of PBHs is expected.

For each PBH, the velocity is drawn randomly from a Maxwell-Boltzmann distribution. The characteristic 
velocity of the distribution is position-dependent. 
The velocity distribution at a given radius is a crucial ingredient, because the accretion rate scales as 
$v^{-3}$, eq.~(\ref{eq:accretion}). 
In order to derive such a distribution, we consider the recent state-of-the-art model for the mass distribution in 
the Milky Way described in~\cite{McMillan2016}, where 6 axis-symmetric components are taken into account 
(bulge, DM halo, thin and thick stellar discs, and HI and molecular gas discs). We then assume that the velocity
 distribution at a distance $R$ from the GC is a Maxwell-Boltzmann with $v_{\rm mean} = v_{\rm circ}(R) = \sqrt{\left( G \, M(<R) / R \right)}$. 
Under the assumption of isotropic orbits\footnote{We verified that, in the high-resolution Aquarius N-body simulations, 
the anisotropy parameter $\beta = \ 1 - \sigma_t/\sigma_r$ is consistent with 0 in the whole range of radii we are interested in, 
therefore the assumption of isotropic orbits is solid.}, an exact computation of the phase-space density could be performed
 by means of the Eddington formalism \cite{Eddington:1915}, as done e.g. in \cite{Fornasa:2014}. We checked 
 that our simple approach is equivalent in the low-velocity tail, up to $v \simeq 40$ km/s. \footnote{M. Fornasa, private communication.} 
 Since our results depend only on PBHs with velocities
 $\lesssim 10$ km/s (see below), we can safely neglect the high-velocity tail and adopt the simple formalism described above.

Given the mass, position and velocity of each PBH (and the gas density), we compute accretion rate, X-ray, and radio 
emission adopting the prescriptions discussed in the previous section.

\begin{figure}
	\includegraphics[width=\columnwidth]{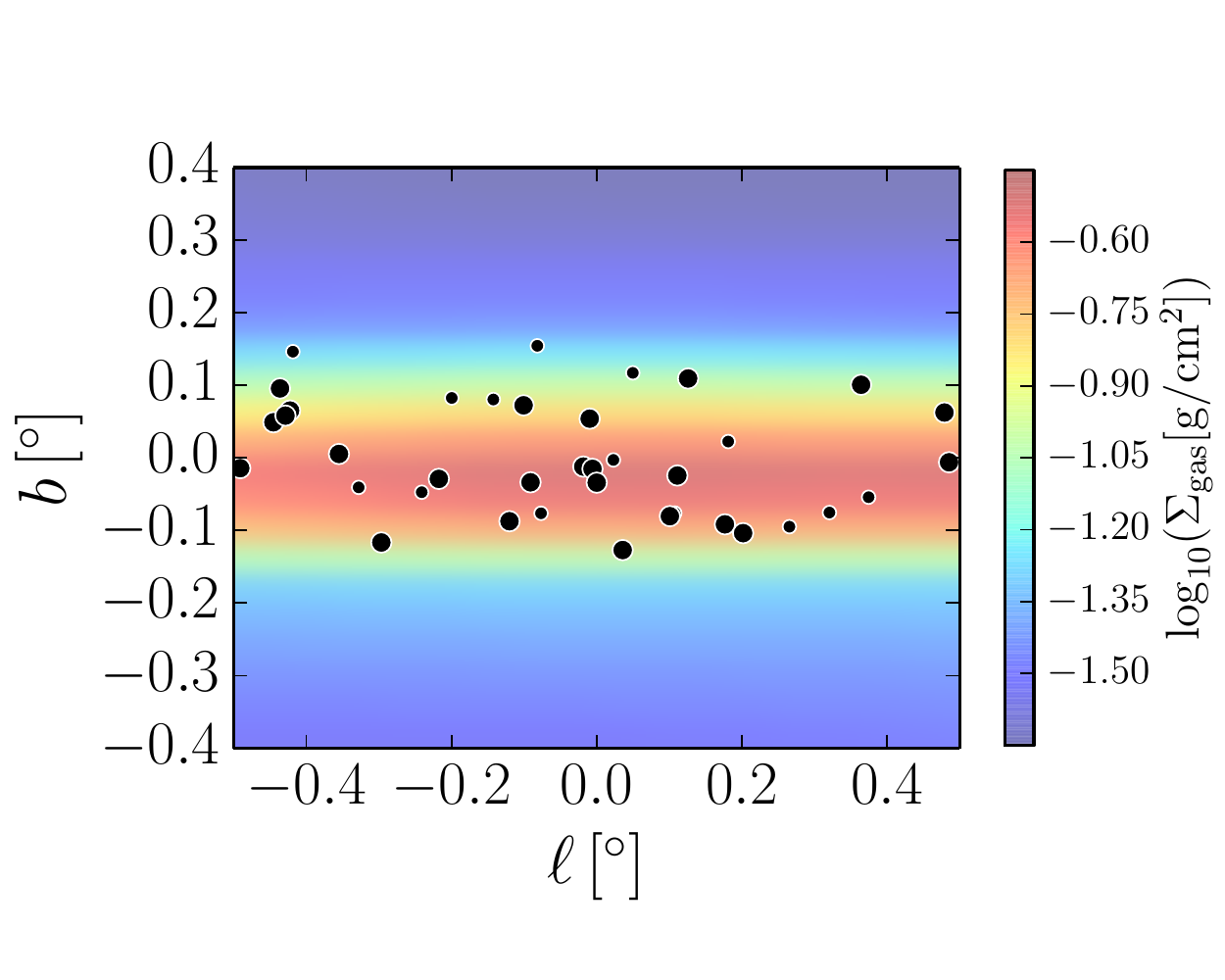}
    \caption{Example of the distribution {  of $30$ $M_{\odot}$ PBHs} detectable by VLA
    in the ROI, for one Monte Carlo realization. 
    The colored background depicts the column gas density. 
    The size of the black points is proportional to the PBH velocity in the range $0.3 - 3$ km/s
    (for detectable PBHs).}
    \label{fig:sources}
\end{figure}

\header{Radio BH candidates} 
The $1.4$~GHz source catalog from a VLA survey of the GC region~\cite{Lazio2008} contains $170$ sources in a $1^{\circ}\times 1^{\circ}$ region centered on the GC. The minimum detectable flux for this catalog is $\sim1$~mJy.

In order to compare our predictions to the observations, we carry out a data analysis on the VLA catalog and check if there can be any BH candidate among the detected sources. 

If any of these sources are accreting BHs, their X-ray and radio emissions should be co-located. We therefore compare the radio catalog with the X-ray point source catalog from \cite{Muno2009}, which contains $9017$ sources detected by \textit{Chandra} in the $0.5-8$~keV band in a $2^{\circ}\times0.8^{\circ}$ band centered on the GC, and search for all sources in both catalogs that have positions within $10''$ of each other.\footnote{This is a very conservative separation. The positional accuracy of \textit{Chandra} is $<1''$. For the VLA, the positional accuracy is typically a small fraction of the synthesized beam, $2''.4\times1''.3$ for the survey in \cite{Lazio2008}{ , taken in A configuration}. A separation of $10''$ is chosen in \cite{Lazio2008} to search for positional coincidences in other radio catalogs; we therefore also adopt $10''$ as the maximum allowed separation.}

We find $24$ sources in both the X-ray and radio catalogs within $10''$ of each other.  
If we assume that these sources are accreting BHs, then their X-ray and radio fluxes should lie on the FP, as explained above. So, we use the FP ({  considering masses from 10 to 100 $M_{\odot}$)} to predict the X-ray flux from the radio flux of each of these objects ($24$ in the very conservative case, $9$ if we exclude likely foreground sources). 

We find that the predicted X-ray fluxes are substantially larger ($\sim3-7$ orders of magnitude) than the flux reported in the catalog from \cite{Muno2009} {  in the whole mass range we consider}. 
We therefore conclude that none of the 24 (or 9 likely Galactic) VLA sources with overlapping positions lie on the FP, and therefore, given the assumptions described above regarding the presence of a jet, we have no BH candidate in our sample. 

\header{X-ray BH candidates}
Hard X-ray emission ($>10$~keV) suffers from far less Galactic absorption than soft X-ray 
emission and is therefore a good band to search for emission from accreting BHs. 

{
We consider sources in the Chandra catalog~\cite{Muno2009} in the $0.5 - 8$ keV band, and those detected by NuSTAR in the $10 - 40$ keV band~\cite{Harrison:2013}. For Chandra (NuSTAR), we consider a small region-of-interest (ROI) including the high-density region of the Galactic Ridge: $-0.9^{\circ} < l < 0.7^{\circ};  -0.3^{\circ} < b < 0.3^{\circ}$ ($-0.9^{\circ} < l < 0.3^{\circ};  -0.1^{\circ} < b < 0.4^{\circ}$).
There are $483$ likely Galactic X-ray sources in the Chandra catalog above a flux threshold of $2\times10^{-6}$~ph~cm$^{-2}$~s$^{-1}$\footnote{"Likely Galactic" sources are defined in \cite{Muno2009} based on their hardness ratios. The exposure across the Chandra survey region is variable and the flux threshold used here is a compromise between maximizing the ROI and the completeness, per \cite{Muno2009}.}, and $70$ NuSTAR sources. 
Since in all cases the corresponding radio flux predicted with the FP would be $3-7$ orders of magnitude below the detection threshold of the VLA survey in \cite{Lazio2008}, we cannot draw any conclusions on the nature of these X-ray sources. Therefore, we consider all of them in our analysis as potential BH candidates (we only remove $\sim 40\%$ of the detected NuSTAR sources that are thought to be cataclysmic variables \cite{Hong:2016}).}



\header{Results}
The main result of the Letter is presented in fig.~\ref{fig:radioBound}.
We display the 2$\sigma$, 3$\sigma$, and 5$\sigma$  constraints on the DM fraction as a function of the PBH mass. 

The upper limits are derived as follows. We perform ${\cal O}(100)$ Monte Carlo simulations for $10$ reference
 values of the mass in the $10 - 100$ $ M_{\odot}$ interval, assuming a DM fraction $f_{\rm DM} = 1$. We 
 determine the mean and standard deviation of the distributions of the predicted number of PBHs with radio fluxes 
 above the VLA threshold and with X-ray fluxes exceeding the { Chandra (NuSTAR)} threshold, in the corresponding ROIs. 
 We verify that the number of bright PBHs is compatible with Poisson statistic and the average predicted number scales linearly 
 with $f_{\rm DM}$. We derive the radio and X-ray bounds by comparing the number of predicted PBHs with the number of 
 BH candidates derived from the analysis of radio and X-ray catalogs described in the previous section.  { For the X-ray bound, we show the result obtained with the more sensitive Chandra catalog. The NuSTAR bound is slightly weaker: It allows us to exclude at $2\sigma$ values of $f_{\rm DM}$ as low as $0.4$ (for $30 \,M_{\odot}$).}

In fig.~\ref{fig:sources}, we show the PBHs detectable by VLA at $1.4$ GHz assuming a PBH mass of 30$M_\odot$ and
DM fraction equal to 1, for one specific Monte Carlo realization. This scenario predicts, on average, { 40 $\pm$ 6} sources above the 
VLA flux threshold  for $30 \, M_\odot$ and, thus, it is excluded by more than { 5$\sigma$} from radio observations. 
However, it is important to understand where the constraining power comes from: The PBHs above the
detection threshold, and thus the ones with the larger X-ray flux, lie in the very
inner region of the Galaxy where the column gas density is the highest and show very small velocities,
in the range $\sim$ 0.3 $-$ 3 km/s. Therefore, the constraints arise from the very low velocity tail of
the distribution and from regions correlated with very high column densities, e.g.~CMZ, as already mentioned above.

\header{Discussion and conclusions}
In this Letter we derive new, strong constraints on the hypothesis that PBHs comprise all of the DM in the Universe. 
In particular, we find that PBHs with $M \simeq 30 M_\odot$, that could be responsible for the gravitational waves detected by LIGO,
contribute less than 20\% to the whole DM density.

In the mass window $10 - 100$ $M_\odot$, our constraints are 
competitive with (and even stronger than) those arising from the study of microlensing events with the MACHO project~\cite{Alcock:2001} (for $\gtrsim 15 M_\odot$) and with those from halo wide binaries~\cite{2009MNRAS.396L..11Q,2014ApJ...790..159M} (for $\gtrsim 60 M_\odot$).
For $M \gtrsim 10 M_\odot$, they are also comparable or stronger than the constraints from the survival of central star clusters in faint dwarf galaxies, in particular in Eridanus II~\cite{Brandt:2016aco,Li:2016utv}.
{Even more stringent constraints arise in principle from the analysis of the Cosmic 
Microwave Background (CMB)~\cite{Ricotti:2007au}. However, 
those arising from the analysis of spectral distortions (based on FIRAS data) turned out to be much weaker
than originally thought~\cite{ClesseBellido2016}, {while the ones} based on the study of CMB anisotropies (see also the recent results 
by~\cite{Chen2016}), are based on assumptions on the accretion of gas on PBHs in the early Universe that are still under debate, as the modeling of the accretion process is based on theoretical arguments, and not directly supported by observations as in our case  (see also the discussion in~\cite{ClesseBellido2016}).

In contrast with~\cite{Ricotti:2007au}, in fact, we adopt a very conservative prescription, 
compatible with current astronomical observations, for both the accretion rate and the radiative efficiency,  
setting the ratio of the actual accretion rate to the Bondi rate, $\lambda$, equal to 0.02}. {  We remark that $\lambda$ probably follows some distribution and is also likely degenerate with $c_S$ and $v_{BH}$---future studies are required to disentangle these.}
Moreover, we exploit for the {first time} in this context the empirical FP relation between radio and X-ray emission, which has 
been observed on a wide class of sources in a large mass range, from X-ray binaries to active Galactic nuclei. By adopting such a relation, 
we are able to predict the expected radio and X-ray luminosities of a population of PBHs in the Galaxy compatible with the DM phase-space distribution, 
as well as to look for BHs candidates in radio and X-ray catalogs.  
We set upper limits on the DM PBH fraction using both radio (VLA) and X-ray { (Chandra, NuSTAR)} point-like source catalogs, 
by comparing the number of expected PBHs above observational thresholds and the observed number of BH candidates in a very narrow region about the GC.

These bounds are robust with respect to the modeling of the full
velocity distribution, since the predicted number of bright PBHs only depends
on the very low-velocity tail ($<$ 10 km/s) where we checked the agreement among different numerical/analytical methods.
Moreover, our limits are independent of the details of the gas distribution (we checked that the bound is still present even with a naive modeling of the CMZ as a sphere with uniform density compatible with the mass constraints provided in \cite{Ferriere2007}).
{  They are also not affected significantly by a shallower DM profile as proposed e.g. in \cite{Calore2015}; however, assuming an even flatter profile like the Burkert one (an extremely conservative assumption for our Galaxy), the bound is present only under the assumption of Bondi accretion}.

We recall that our limits hold for a narrow mass function; a detailed study of the impact of different mass distributions is beyond the scope of the present paper and postponed to a future work.

{  Although our radio and X-ray bounds vanish for $\lambda~\lesssim~~10^{-2}$,}
future instruments will be able to verify better  the accretion model as well as the PBH DM fraction.
In particular, given the significant increase in sensitivity of future radio telescopes, 
we expect an important part of the yet-allowed parameter space to be probed by upcoming facilities
such as MeerKAT and, later, SKA.
Using the radiometer equation~\citep{1984bens.work..234D}, the minimum 
(1$\sigma$) detectable radio flux is $S_{\nu, \rm rms} = (T_{\rm sky} + T_{\rm rx})/(G \, \sqrt{2 \, T_{\rm obs} \, \Delta\nu })$.
{ For SKA1-MID (1.4 GHz), we assume gain G = 15 K/Jy, receiver temperature $T_{\rm rx}$ = 25 K, sky temperature towards the GC 
$T_{\rm sky}$ = 70 K, and bandwidth $\Delta\nu$ = 770 MHz~\cite{Calore:2015bsx}.}

{  For one-hour exposure, the instrumental detection sensitivity of SKA1-MID turns out to be $\sim 2.7$ $\mu$Jy (significantly above the source confusion limit), which would give O($2000$) detectable PBHs for our reference value $\lambda = 0.02$ ($f_{\rm DM}$= 1, and $M = 30 M_\odot$). 

SKA will therefore be able to either place a very strong constraint in absence of BH candidate detection, or detect a subdominant population of PBHs (although the expected population of astrophysical BHs becomes comparable with the primordial one for DM fractions lower than $\sim 10^{-3}$).
With an even longer exposure ($\simeq 1000$ h, 85 nJy sensitivity), achievable for dedicated deep field continuum observations, such strong constraints can be placed by SKA even under the assumption of extremely low values of $\lambda$, ${\cal O}(10^{-3})$.}

{Interestingly, our procedure can also be applied {  in order to extend work on searches for astrophysical BHs in the Galaxy \cite{Agol2002,Maccarone2005MNRAS,Fender:2013}}, adopting the realistic spatial and velocity distributions expected for those objects. Our formalism has the potential to characterize this guaranteed population of objects in future analyses.} 

\vskip 5mm

\header{Acknowledgments} We are indebted to M. Fornasa for verifying of the phase-space velocity at low radii with the Eddington formalism. We thank P. D. Serpico for his useful comments on the manuscript, and D. Baumann, P. Crumley, B. Freivogel, A. King, A. Urbano, J. Vink, and C. Weniger for useful discussions. Finally, we thank the anonymous reviewers for their insightful comments and suggestions.

\bibliography{sample} 

\end{document}